\documentclass[aps,twocolumn,prb,showpacs,amsmath,amssymb]{revtex4-1}
\usepackage[colorlinks,citecolor=blue,linkcolor=cyan]{hyperref}
\usepackage[usenames,dvipsnames,svgnames,table]{xcolor}

\usepackage{dcolumn}
\usepackage{bm}
\usepackage{graphicx}
\usepackage{epstopdf}
\usepackage{color}
\usepackage{placeins}
\usepackage{braket}

\usepackage[usenames,dvipsnames,svgnames]{xcolor}

\begin{document}

\newcommand{\bfk}{{\bf k}}
\newcommand{\bfR}{{\bf R}}
\newcommand{\Hamil}{{\cal H}}

\newcommand{\vc}[1]{\mathbf{#1}}
\newcommand{\e}[1]{\mathrm{e}^{#1}}
\newcommand{\I}{\mathrm{i}}

\title{From chiral $d$-wave to nodal line superconductivity in the harmonic honeycomb lattices}
\author{Johann Schmidt}
 \affiliation{Department of Physics and Astronomy, Uppsala University, Box 516, S-751 20 Uppsala, Sweden}
 \author{Adrien Bouhon}
 \affiliation{Department of Physics and Astronomy, Uppsala University, Box 516, S-751 20 Uppsala, Sweden}
 \author{Annica M. Black-Schaffer}
  \affiliation{Department of Physics and Astronomy, Uppsala University, Box 516, S-751 20 Uppsala, Sweden}
\date{\today}

\begin{abstract}
Motivated by the recent realization of the three-dimensional hyperhoneycomb and stripyhoneycomb lattices in lithium iridate (Li$_2$IrO$_3$), we study the possible spin-singlet superconducting states on the whole series of harmonic honeycomb lattices. Beginning with an isolated out-of-plane twist making the honeycomb lattice three-dimensional, we find that the chiral $d\pm \I d'$ state, well-known from the honeycomb lattice, is realized in the largest members of the series at low to intermediate doping. Along the twist, four chiral edge states form a two-dimensional dispersive band. Reducing the distance between the twists to form the smaller members of the harmonic honeycomb lattices, the degeneracy between the $d$-wave states is lifted, which finally destroys the chiral state. By analyzing the hyper- and stripyhoneycomb lattices and generalizing using the $D_{2h}$ point group of all the harmonic honeycomb lattices, we show that the superconducting state often belongs to the trivial irreducible representation. This state has nodal lines at low to intermediate doping, which is possible because the full lattice symmetry allows sign changes between different sets of bonds. We also find time-reversal symmetry broken states, which are either fully gapped or feature nodal points, in certain parts of the phase diagram.
Finally, we draw a comparison between the states classified in terms of the $D_{2h}$ symmetries and those observed on the $D_{6h}$ honeycomb lattice.

\end{abstract}
\maketitle

%
\section{Introduction}
Honeycomb materials offer an attractive path to the realization of a chiral spin-singlet superconducting state. The sixfold rotation symmetry of the honeycomb lattice makes the two $d$-wave solutions, favored for electron-driven superconductivity in proximity to an antiferromagnetic state, degenerate, and thus a time-reversal symmetry breaking and fully gapped $d_{x^2-y^2}\pm i d_{xy}$-wave state (or short $d\pm id'$) is very generally preferred.\cite{BlackSchafferHonerkamp2014}
This state is characterized by the phase of the order parameter winding twice around the Brillouin zone center with the chirality defined by the sign of the imaginary part, which functions as a definition of a topological invariant.\cite{Schnyder2008, BlackSchafferHonerkamp2014} Due to the bulk-boundary correspondence, open boundaries necessarily host a pair of co-propagating, or chiral, edge states.\cite{Volovik1997, BlackSchaffer2012}

One natural material to realize the $d+id'$-wave state is graphene.\cite{Novoselov2006} Both largely phenomenological models\cite{BlackSchaffer2007, Honerkamp2008, Pathak2010} and different renormalization group (RG) techniques\cite{Nandkishore2012, Wang2012, Kiesel2012} have found a $d+id'$ superconducting state in doped graphene, in particular close to the van Hove singularity at quarter filling. In recent years the layered honeycomb lithium- and sodium iridates, Li$_2$IrO$_3$ and Na$_2$IrO$_3$, have also generated significant experimental and theoretical interest.\cite{Jackeli2009, Chaloupka2010, Singh2010, Singh2012} In these materials the lattice geometry combines with strong spin-orbit coupling and strong correlations to possibly generate a quantum spin liquid state with Majorana excitations interesting for quantum computation\cite{Kitaev2006} in the undoped state. From the viewpoint of superconductivity, theoretical studies have found that doping these materials can also give a spin-triplet $p$-wave state, possibly topologically non-trivial,\cite{Hyart2012,You2012} competing with the spin-singlet chiral $d$-wave state.\cite{Okamoto2013, Scherer2014} A finite momentum pairing state has also been discussed.\cite{Liu2015}

Very recently, lithium iridate has also been synthesized in two other crystal structures which have been dubbed hyperhoneycomb~\cite{Takayama2015} and stripyhoneycomb.\cite{Modic2014} These are envisioned to be the two smallest members of a series of 3D lattices called the harmonic honeycomb lattices, which can be viewed as regions of honeycomb lattice twisted against each other to form a truly 3D structure. In the past few years, there have been many experimental and theoretical studies of these materials in the context of magnetism, with the hope of finding a quantum spin liquid ground state.\cite{Mandal2009, Lee2014, Kimchi2014, Biffin2014-hyper,Takayama2015,Biffin2014-stripy,Kimchi2015, Lee2015} Other proposals of exotic physics on these lattices include a Weyl spin liquid\cite{Herrmanns2015} and line nodes.\cite{Ezawa2016}

Despite the large interest in the strongly correlated ground state of the undoped harmonic honeycomb materials, the possibility for superconductivity upon doping has so far not been explored. Because the larger members of the harmonic honeycomb series consist of large honeycomb regions, the chiral $d\pm id'$-wave symmetry should realistically be very competitive for any spin-singlet superconducting state.
However, all harmonic honeycomb lattices globally break the sixfold rotation symmetry and thus the intrinsic 3D nature should also significantly influence superconductivity, which makes the resulting state hard to predict. 

In this work we study the possible spin-singlet superconducting states in the doped harmonic honeycomb lattice series. We start with investigating an isolated twist separating two large honeycomb regions. Indeed, we find that the $d\pm\I d'$ state is stabilized on both sides of the twist for all but very high doping levels. The relative chirality between both sides is not constrained by the twist and each twist host a total of four chiral edge states. The twist thus largely acts as an effective open boundary although we find that it allows for hybridization between the chiral edge states, such that they form 2D dispersive states in the twist plane.
Adding twists periodically and subsequently systematically reducing the size of the honeycomb regions, we observe pair breaking  in the $d_{xy}$ channel. This leads to the destruction of the chiral $d\pm id'$-wave state below a critical size of the harmonic honeycomb lattice. The remaining order parameter does not completely coincide with the $d_{x^2-y^2}$ state either, since sixfold rotation symmetry is not fully preserved. Studying the smallest members of the harmonic honeycomb lattice series, the hyper- and stripyhoneycomb lattice, we discover that the superconducting order parameter instead forms a nodal state belonging to the trivial representation. The formation of nodal lines is possible even in the trivial representation because the full lattice symmetry allows for sign changes between certain set of bonds. Other orders found in certain parts of the phase diagram include time-reversal symmetry broken states, which are either fully gapped or feature nodal points. Very large doping always leads to a fully gapped extended $s$-wave state throughout the whole harmonic honeycomb series. 

This article is structured in the following way. In Section~\ref{sec:lattice} we build up the harmonic honeycomb lattices from a simple starting point - the twist. We then introduce the model in Section~\ref{sec:model}. In Section~\ref{sec:twist}, we study an isolated twist and investigate the properties of the twist edge states. The pair breaking effect of a periodic structure with varying distances between twists is investigated in Section~\ref{sec:interm}. We then perform a symmetry analysis of the smallest members of the harmonic honeycomb series in Section~\ref{sec:hypstrip} and calculate their superconducting phase diagrams. Finally, in Section~\ref{sec:connection}, we build upon the observed behaviors and draw conclusions for all members of the harmonic honeycombs, before we summarize our results in Section~\ref{sec:conclusion}.

%
\section{Harmonic honeycomb lattices\label{sec:lattice}}
\begin{figure*}[t]
    \includegraphics{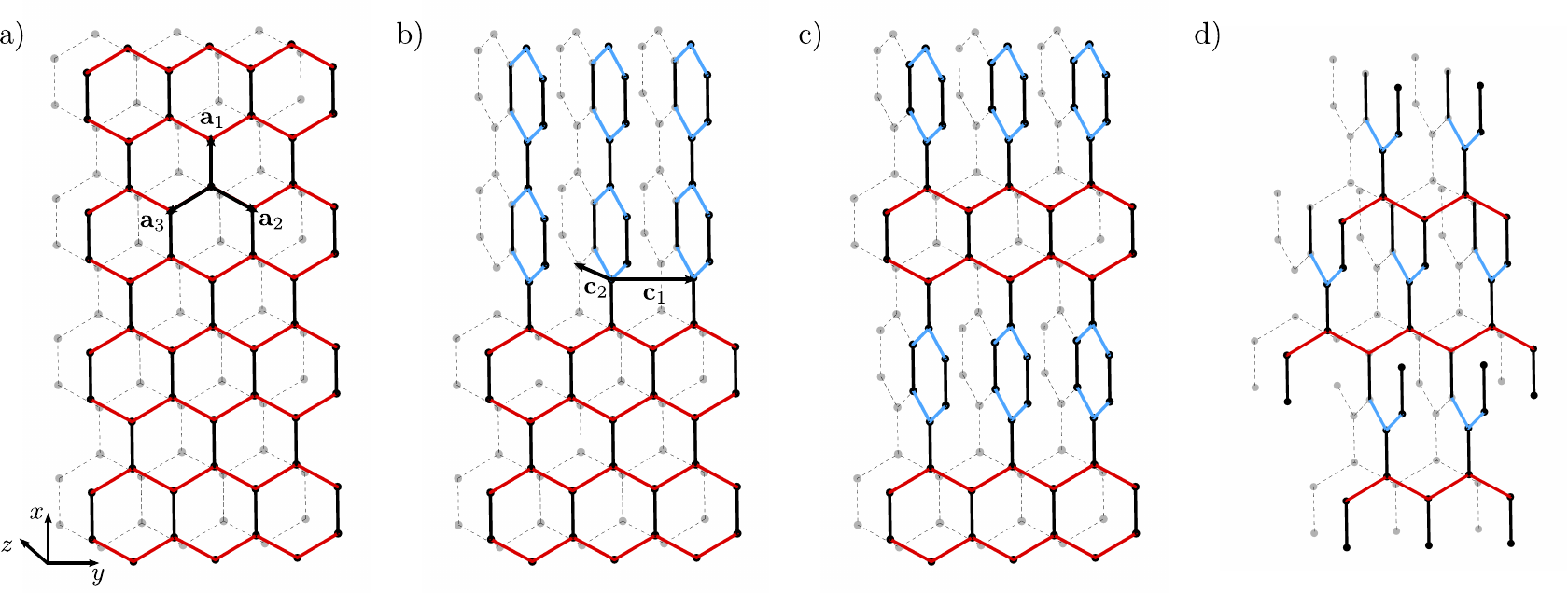}
    \caption{Harmonic honeycomb lattice series. (a): Stack of 2D honeycomb sheets with the three nearest neighbor vectors $\vc{a}_1$,$\vc{a}_2$ and $\vc{a}_3$. (b): When the zigzag nearest neighbor vectors change from $\vc{a}_i$ to $\vc{a}_i'$, a twist structure is created. The new lattice vectors for the twist $\vc{c}_1$ and $\vc{c}_2$ are also depicted. (c): The stripyhoneycomb ($\mathcal{H}\braket{1}$) and (d): hyperhoneycomb ($\mathcal{H}\braket{0}$) lattices are the two smallest members of the series. In all panels the zigzag bonds $\vc{a}_2$ and $\vc{a}_3$ are drawn in red, rotated zigzag bonds $\vc{a}'_2$ and $\vc{a}'_3$ in drawn in blue, and additional honeycomb layers are marked in grey. Image inspired by Modic {\it et al.}\cite{Modic2014}     \label{fig:lattices}}
\end{figure*}

The harmonic honeycomb lattices are tri-coordinated 3D lattices that are related to the honeycomb lattice.
The honeycomb lattice has the nearest neighbor vectors $\vc{a}_1 = \left(1,0,0\right)$, $\vc{a}_2 = \frac{1}{2}\left(-1, \sqrt{3},0\right)$ and $\vc{a}_3 = \frac{1}{2}\left(-1,-\sqrt{3},0\right)$, in units of the nearest neighbor distance $a$ (which we will use as the unit of length) and as shown in Fig.~\ref{fig:lattices}(a).
In order to produce the harmonic honeycomb lattices, some of the zigzag bonds $\vc{a}_2$ and $\vc{a}_3$ (red bonds in Fig.~\ref{fig:lattices}) are rotated out of the $x-y$ plane by an angle $\alpha$ to instead yield $\vc{a}'_2 = \frac{1}{2}\left(-1, \sqrt{3} \cos(\alpha), \sqrt{3} \sin(\alpha)\right)$, $\vc{a}'_3 = \frac{1}{2}\left(-1,-\sqrt{3} \cos(\alpha), -\sqrt{3} \sin(\alpha)\right)$ (blue bonds in Fig.~\ref{fig:lattices}). The straight nearest neighbor vector $\vc{a}_1$ (black bonds in Fig.~\ref{fig:lattices}) is still the same throughout the lattice. 
The switch from the unprimed to the primed nearest neighbor vectors occurs at a fixed $x$-position and parallel to the $y$ and $z$-directions. We call this lattice change a ``twist" as is depicted in Fig.~\ref{fig:lattices}(b).

At a twist, a plane spanned by the $\vc{a}_2$ and $\vc{a}_3$ vectors is connected to infinitely many parallel planes spanned by the primed zigzag vectors and vice versa.
Thus the rotation of $\vc{a}'_2$ and $\vc{a}'_3$ out of the $x-y$ plane creates a fully 3D lattice.
The resulting twist structure always has a periodicity in the $y-z$ plane with lattice vectors $\vc{c}_1 = \sqrt{3} \left(0,1,0\right)$ and $\vc{c}_2 = \sqrt{3} \left(0,\cos(\alpha),\sin(\alpha) \right)$, as demonstrated in Fig.~\ref{fig:lattices}(b). The reciprocal lattice vectors for this plane are $\vc{b}_1 = \frac{2 \pi}{\sqrt{3}} \left(0,1,-\cot(\alpha)\right)$ and $\vc{b}_2 = \frac{2 \pi}{\sqrt{3}} \left(0,0,\csc(\alpha)\right)$, resulting in a first Brillouin zone depicted in the inset in Fig.~\ref{fig:isolatedtwistspectrum}.

The twist can also be periodically repeated at different $x$ positions, creating an infinite amount of possible lattice structures -- the harmonic honeycomb lattice series. When the regions between twists consist of $n$ completed hexagons the resulting lattice is labeled $\mathcal{H}\braket{n}$.\cite{Modic2014} Figures~\ref{fig:lattices}(c) and (d) depict the two smallest members of this family: the stripyhoneycomb ($\mathcal{H}\braket{1}$) and hyperhoneycomb ($\mathcal{H}\braket{0}$) lattices. Both of these structures have very recently been stabilized experimentally in lithium iridate Li$_2$IrO$_3$.\cite{Modic2014,Takayama2015} In these compounds the harmonic honeycomb lattice is formed by the iridium ions, which are surrounded by oxygen octahedra. This arrangement leads to an angle of rotation $\alpha = \arccos\left(\frac{1}{3}\right)\approx 70^\circ$,\cite{Modic2014} which we adopt throughout this paper. Another possibility would be the very recent proposed realization of the harmonic honeycomb lattices involving carbon atoms with $\alpha = 90^\circ$.\cite{Mullen2015}

%
\section{Model for superconductivity \label{sec:model}}
We are here interested in the superconducting state which might occur upon doping a strongly correlated harmonic honeycomb material, having the already proposed spin-liquids Li$_2$IrO$_3$ in mind. In order to most clearly elucidate the physics of the possible superconducting states, we consider the simplest possible kinetic energy term in the harmonic honeycomb lattices appearing upon doping, consisting of a nearest neighbor hopping $t$ and chemical potential $\mu$
\begin{align}
H_0 = -t \sum_{\braket{i,j},\sigma} \left( c^\dagger_{j\sigma} c_{i\sigma} + {\rm H.c.} \right)
+ \mu \sum_{i,\sigma} c^\dagger_{i\sigma} c_{i\sigma},
\end{align}
where $c_{i\sigma}$ is the annihilation operator on site $i$ with spin $\sigma$.
We further consider superconductivity in the spin-singlet channel, where we capture all relevant spatial pairing symmetries by using a nearest neighbor bond pairing order parameter $\Delta_{ij}$, such that the total Hamiltonian reads
\begin{align}
H = H_0 + \sum_{\braket{i,j}} \Delta_{ij} \left( c^\dagger_{i\uparrow}c^\dagger_{j\downarrow} - c^\dagger_{i\downarrow} c^\dagger_{j\uparrow} \right) + {\rm H.c.} + \frac{2 |\Delta_{ij}|^2}{J}.
\end{align}
Exactly such a superconducting term is generated upon doping a material with an antiferromagnetic nearest neighbor Heisenberg interaction $J$,\cite{Anderson1973, Baskaran1987, BlackSchaffer2007}, which has been found to be significant in the known harmonic honeycomb materials.\cite{Kimchi2015,Kim2015-hyper,Katukuri2016} We leave the study of the other interaction terms discussed in connection with the iridates, such as the Kitaev interaction,\cite{Jackeli2009, Chaloupka2010, Lee2014, Kimchi2014} to future work. An inclusion of such a term will introduce additional spin-triplet pairing.\cite{Hyart2012, You2012}
The superconducting order parameter is determined through the self-consistency equation
\begin{align}
\Delta_{ij} &= -\frac{J}{2} \braket{c_{i\downarrow} c_{j\uparrow} - c_{i\uparrow} c_{j\downarrow}}.
\end{align}
In order to capture all possible superconducting states we solve self-consistently for the superconducting order parameter independently on each bond, with no presumed symmetry relations inbetween different bonds.

The model discussed here has been studied before on the honeycomb lattice, which has a $D_{6h}$ point group symmetry.\cite{BlackSchaffer2007} In that case, there are three bond order parameters $\Delta_{ij}$, one for each nearest neighbor bond $\vc{a}_j$. For easy treatment, they can be arranged into a vector order parameter $\vc{\Delta} = \left(\Delta_1,\Delta_2,\Delta_3 \right)$. With three different bond orders there exist in total three different solutions: Two $d$-wave order parameters, belonging to the 2D irreducible representation $E_{2g}$, and an extended $s$-wave state, which belongs to the trivial representation $A_{1g}$.\cite{BlackSchaffer2007, BlackSchafferHonerkamp2014} 
The labeling of the states refers to their behavior under the symmetry transformations of the $D_{6h}$ point group, and is the symmetry of the intraband order parameter across the whole Brillouin zone. 
These states can also be classified by their basis functions in real space, $\vc{\Delta}_{d_{xy}} = \frac{1}{\sqrt{2}}(0,1,-1)$, $\vc{\Delta}_{d_{x^2-y^2}} = \frac{1}{\sqrt{6}}(2,-1,-1)$ and $\vc{\Delta}_{s_{ext}} = \frac{1}{\sqrt{3}}(1,1,1)$. In large parts of the $J-\mu$ phase diagram a linear combination of the two $d$-waves of the form $d_{x^2-y^2} \pm \I d_{xy} = d\pm id'$
is stabilized at all temperatures below $T_c$, with the two different chiralities ($\pm$) being degenerate.\cite{BlackSchaffer2007, Lothman2014} For large doping, beyond the van Hove singularity at $\mu = t$, or for very strong interaction strength $J$, the extended $s$-wave state is instead favored. 
In order to characterize the superconducting solutions for the harmonic honeycomb lattices it is useful to define an overlap of the order parameter as $\frac{|\vc{\Delta} \cdot \vc{\Delta}_x|^2}{|\vc{\Delta}|}$ for the different order parameter symmetries $x$ of the honeycomb lattice.

\section{Isolated honeycomb twists\label{sec:twist}}
In the limit of large distances between twists, the superconducting order parameter of the harmonic honeycombs should be very similar to that of a regular honeycomb lattice. To study this hypothesis, we set up a slab, which is periodic in the $y$- and $z$-direction but has open boundary conditions in the $x$-direction. We use a slab of length $150$ and add a single twist at $x=75$. This is an extended version of the lattice in Fig.~\ref{fig:lattices}(b). To the left of the twist, the zigzag bonds are $\vc{a}_2$ and $\vc{a}_3$, to the right they are $\vc{a}'_2$ and $\vc{a}'_3$. This way the twist connects two stacks of honeycomb lattice with different normal vectors. We characterize the order parameter at each site by the three nearest neighbor bond order parameters around it and calculate the overlap with the three basis functions of the honeycomb lattice.

First, we study if the $d\pm\I d'$ state can be stabilized in the regions on each side of the twist. We therefore fix the parameters to $J=1.1\,t$ and $\mu = 1.0\,t$, which places the system firmly within the $d\pm\I d'$ region (from here on we measure all energies in units of $t$, if not explicitly stated otherwise). These parameters lead to the largest order parameter strength within the $d\pm\I d'$ region.\cite{Lothman2014} This also guarantees a short coherence length for the superconducting state, minimizing the effects of edges~\cite{BlackSchaffer2012} and thus enables us to restrict the calculations to smaller system sizes. We have carefully checked our conclusions with other parameter sets within the $d\pm\I d'$ region, especially for lower chemical potential $\mu$.
 
\begin{figure}[t]
  \includegraphics[width=.4\textwidth]{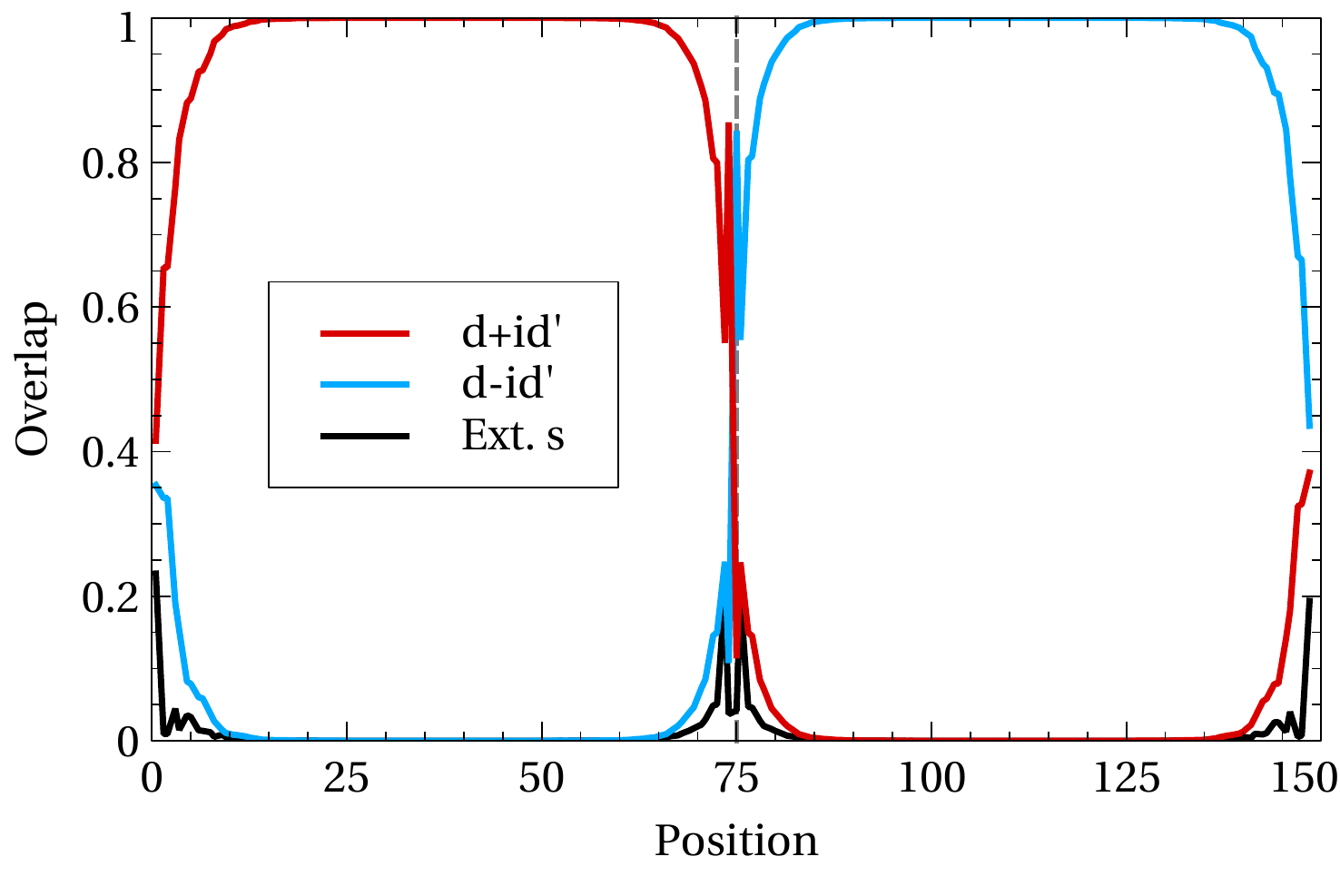}
  \caption{Overlap of the superconducting order parameter with the $d+id'$, $d-id'$, and extended-$s$ basis functions at each atomic site for an isolated twist in the middle of a 150 long slab. Far away from the twist, the results of the honeycomb lattice are recovered and the overlap with the $d\pm\I d'$ states reaches 100\%. Both the twist and the open outer edges lift the degeneracy of the $d$-wave solutions, thus destroying the $d\pm\I d'$ state.
  \label{fig:isolatedtwistoverlap}}
\end{figure}

Figure~\ref{fig:isolatedtwistoverlap} shows the overlap of the calculated order parameter with the honyecomb lattice basis functions as a function of position. The twist is clearly visible in the center of the figure. Far away from the twists, at the center of the regions, the calculated order parameter shows a 100\% overlap with the $d+\I d'$ state to the left of the twist and with the $d-\I d'$ state to the right. Furthermore, the strength of the order parameter in the middle of the regions agrees well with the value of a bulk honeycomb system. Far away from the twist, the system thus clearly behaves like a regular honeycomb lattice.
On the honeycomb lattice, the $d+\I d'$ and $d-\I d'$ states are degenerate. We find no energy difference when changing the chirality on either side of the twist. Thus the chirality of two regions separated by a single twist is completely independent.

Another result that directly carries over from the 2D honeycomb lattice is the suppression of the $d_{xy}$ state on the open outer boundaries of the slab, destroying the $d\pm\I d'$ state.\cite{BlackSchaffer2012} At the same time, the $d_{x^2-y^2}$ state is enhanced and a small overlap with the extended $s$-wave state is also observed near the edges. This has been explained by noting that $\Delta_2 = \Delta_3$ in both the $d_{x^2-y^2}$ and extended $s$-wave states, which respects the translational symmetry of the zigzag edge. On the other hand, in the $d_{xy}$ state $\Delta_2 \neq \Delta_3$, which makes it energetically unfavorable on a zigzag edge and thus the zigzag edge is pair breaking for the $d_{xy}$ state. We find that the order parameter around the twist behaves in a very similar way. The $d\pm\I d'$ state is destroyed near the twist bonds because of the suppression of the $d_{xy}$ state, but it recovers in a fashion very similar to what is observed near the open outer zigzag edges. The order parameter also shows a finite overlap with the extended $s$-wave state at the twist. Moreover, the magnitude of $\Delta$ is reduced to about a quarter of the bulk value on the bond that bridges the twist. 

\begin{figure}[t]
  \includegraphics[width=.4\textwidth]{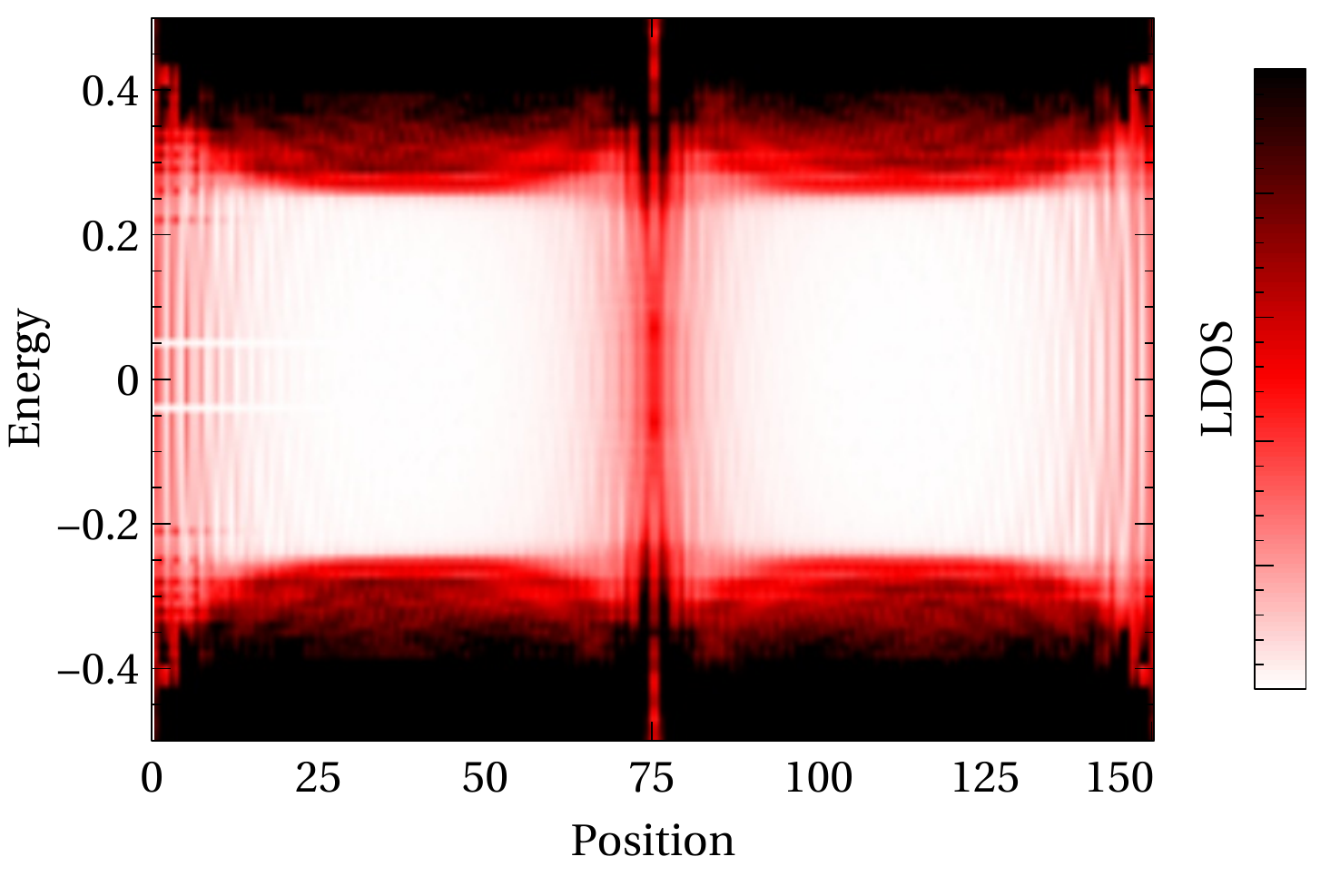}
  \caption{LDOS of an isolated twist in the in the center of a 150 long slab as a function of position. High densities are shown in red and black, while white corresponds to an absence of states. Apart from the edge states near the open outer boundaries, there are also subgap states at the twist.  \label{fig:isolatedtwistLDOS}}
\end{figure}

The $d\pm\I d'$ superconducting state has a Chern number of $\pm 2$, due to the winding of the superconducting order parameter. Thus all edges should host a pair of co-propagating, or chiral, edge states.\cite{Volovik1997, BlackSchaffer2012} Figure~\ref{fig:isolatedtwistLDOS} displays the local density of states (LDOS) as a function of position. In addition to the expected edge states on the open outer boundaries, there are also subgap states visible at the twist. Interestingly, this behavior is observed independently of the chiralities of the two regions separated by the twist. Here we note that the Chern number is only defined in 2D and the twist creates a boundary between two 2D systems with different orientations only connected in isolated points, which motivates why the ground state is not sensitive to the chirality of the states on each side of the twist.

\begin{figure}[t]
  \includegraphics[width=.4\textwidth]{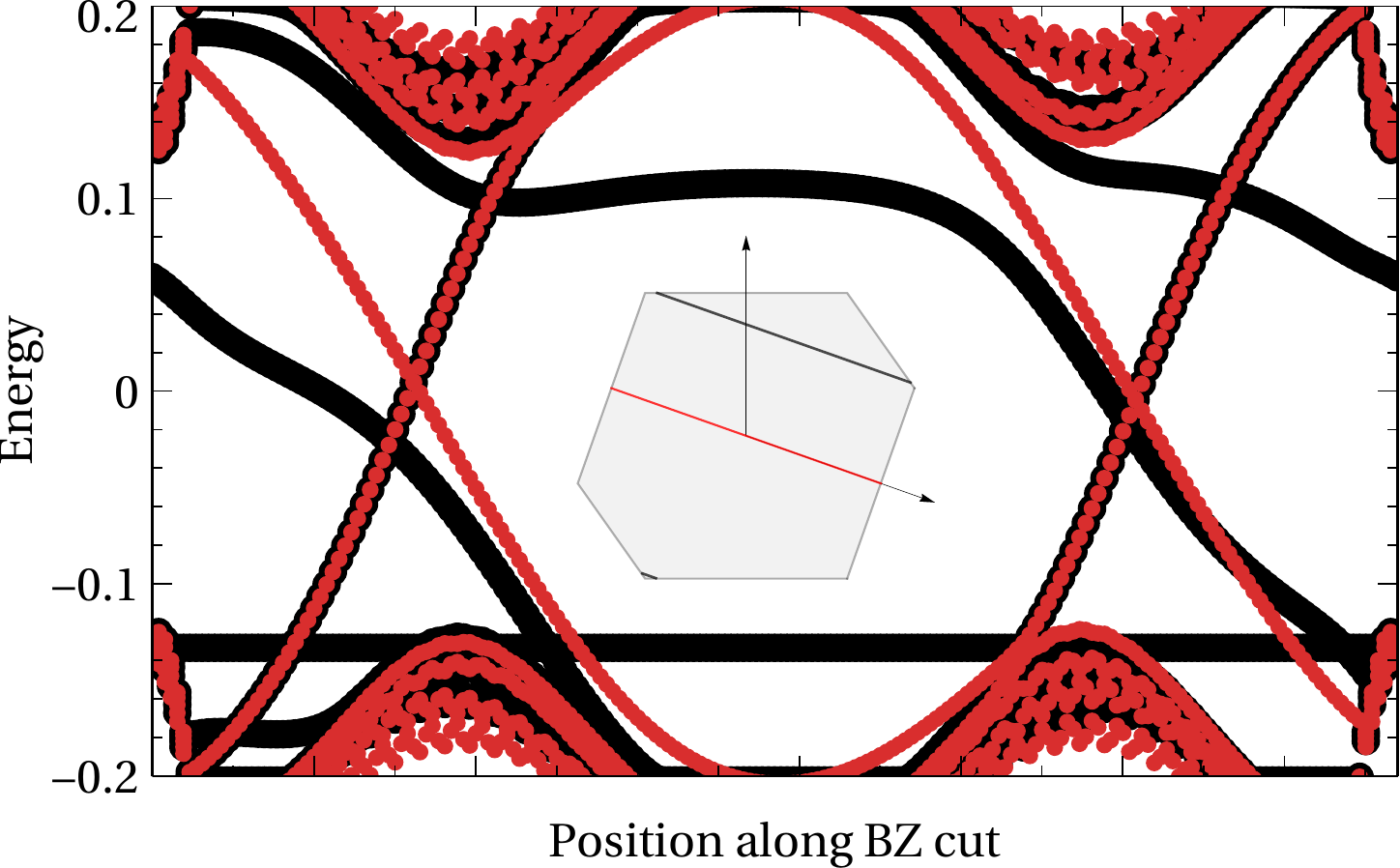}
  \caption{Quasiparticle spectrum plotted along two parallel cuts in the $\vc{b}_1$ direction through the 2D Brillouin zone formed by the twist region (see inset for cuts). In both cases there are four gapless states connecting the conduction and valence band. The edge states along the open outer boundaries do not change between the two cuts. The states at the twist form a 2D dispersive band, which changes between the cuts. \label{fig:isolatedtwistspectrum}}
\end{figure}

In total we expect eight chiral edge states for a slab with a single twist; two at the left edge, four around the twist, and two more at the right edge. Cutting the Brillouin zone in the direction of $\vc{b}_1$ reveals four states crossing the gap  in Fig.~\ref{fig:isolatedtwistspectrum} (red lines). Two of the states do not change when looking at a parallel cut (black lines). This means that they can only disperse along the honeycomb sheets corresponding to the $\vc{b}_1$ direction and must thus be located at the left outer edge of the system. The subgap states at the twist, however, have the possibility to disperse in the directions of both $\vc{b}_1$ and $\vc{b}_2$ and we find that they change between the two cuts. This clearly highlights that the twist edge states hybridize across the twist, forming a 2D band. The two edge states on the right outer edge of the system only disperse in the $\vc{b}_2$ direction. The horizontal line in the spectrum that is seen along the black cut, but not along the red cut, is such an right outer edge state.
In summary, we find that a single twist acts very similar to an outer zigzag edge; the magnitude of the order parameter is suppressed, the twist is heavily pair breaking for the $d_{xy}$-wave order, and the twist carries two chiral edge states on each side of the twist. Notably, the twist does not act as a domain wall, which would enforce a preference on the chirality of the two separated regions. At the same time the twist still allows the four chiral edge states to hybridize such that they form 2D edge bands, which is clearly different from regular 1D zigzag edges.

%
\section{Intermediate harmonic honeycomb lattices \label{sec:interm}}
Next, we introduce periodic boundary conditions also in the $x$-direction and study the superconducting state with changing distance between twists. Once again, we restrict  the parameters to $J=1.1$ and $\mu=1.0$ to obtain the smallest healing length within the $d\pm\I d'$ region of the honeycomb lattice. To further reduce the computational cost, we adjust the Brillouin zone sampling. For distances larger than $25$ between the twists we sample the $k_x$ direction only at the $\Gamma$ point, while we still use a dense sampling for $k_y$ and $k_z$. For smaller distances we use a regular grid in all three dimensions covering the full Brillouin zone. We have carefully checked our results for other parameters and with different Brillouin zone samplings.

\begin{figure}[t]
  \includegraphics[width=.4\textwidth]{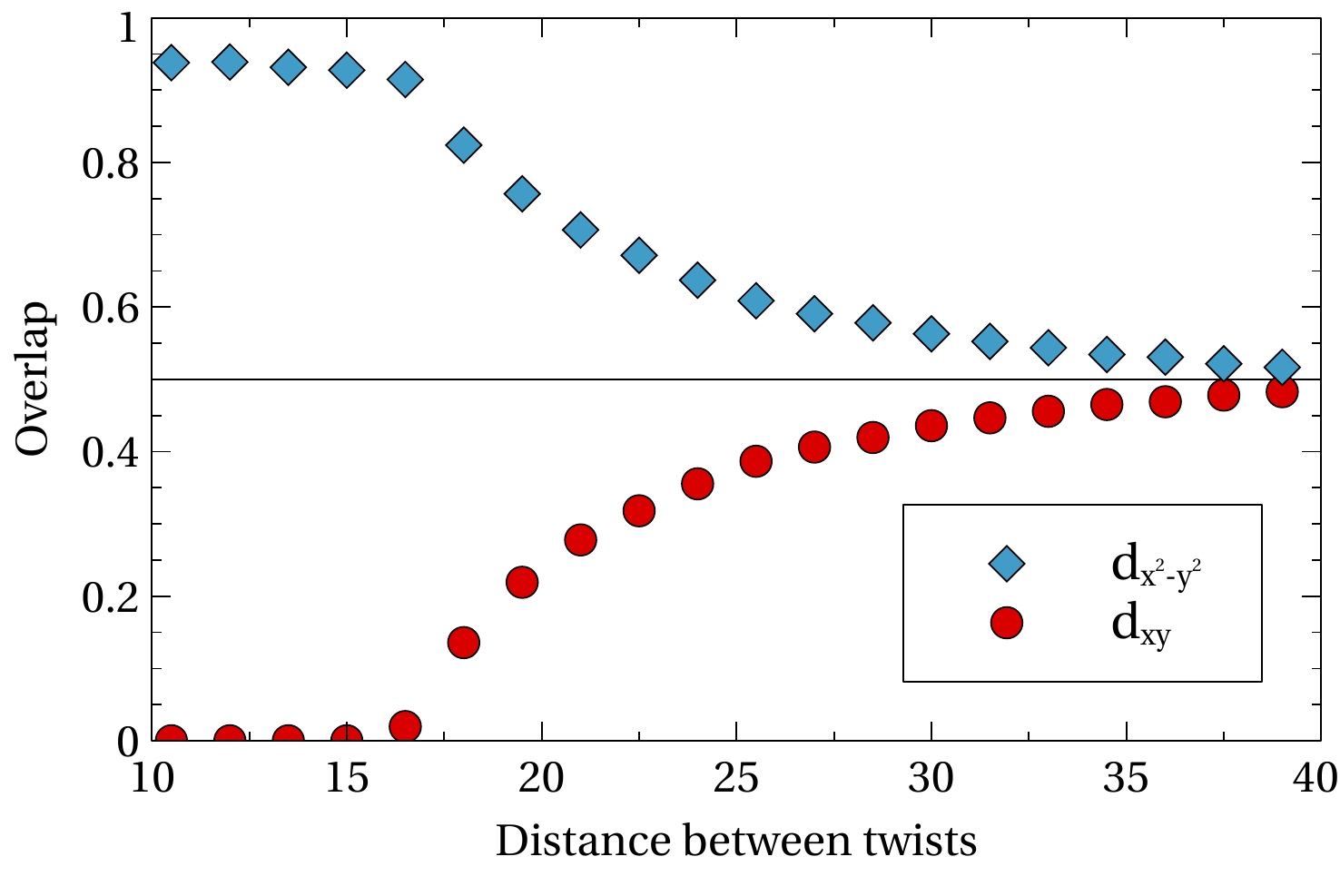}
  \caption{Overlap of the order parameter with the two $d$-wave solutions in the center of the regions between the twists as a function of distance between twists. For large distances the overlap with both $d$-wave states is equal and a perfect $d\pm\I d'$ state is formed. When the distance between twists is decreased, the $d_{xy}$-state is suppressed. \label{fig:dvsL}}
\end{figure}
The influence of the distance between twists is shown in Fig.~\ref{fig:dvsL}, where we plot the overlap of the calculated order parameter with the two $d$-wave solutions at the center of the regions between twists. At large distances, the order parameter has the same overlap with both $d$-wave states and the system is in a perfect $d\pm\I d'$ state. At some smaller distance, where the $d\pm\I d'$ state is still realized, the edge states start to hybridize across the regions, causing the opening of an energy gap in the twist edge states. For even shorter distances, the proximity to the twist suppresses the $d_{xy}$-state also in the center of the regions between twists. The characteristic length scale at which the degeneracy of the $d$-wave states is significantly lifted depends on the coherence length of the superconducting order, which in turn is determined by the strength of the coupling constant and doping. For our set of parameters, which was chosen to give a small healing length at edges, this occurs at a distance of about $25$. For other values of $J$ and $\mu$ within the honeycomb $d \pm \I d'$ region, the degeneracy is lifted at different distances. 
This suppression of the $d_{xy}$-wave symmetry is the same as observed around the isolated twist and near the open zigzag edges. In those cases, however, the system was large enough so that the honeycomb bulk solution was recovered far away from the twist. Now, the regions between the twists are too small to at all stabilize the $d \pm \I d'$ state. 
When the $d_{xy}$ order is completely suppressed, the overlap with the $d_{x^2-y^2}$ order becomes very large. However, Figure~\ref{fig:dvsL} shows that it does not reach unity. The remaining percentage is due to the extended $s$-wave.
When the parameters $J$ and $\mu$ are instead chosen in such a way, that the extended $s$-wave state is stabilized on the 2D honeycomb lattice, the order parameter remains in the extended $s$-wave state throughout the whole harmonic honeycomb series.

Finally, it should be noted that all these classifications are based on the $D_{6h}$ symmetry of the full honeycomb lattice. At small distances between twists, one can no longer assume that this symmetry is fulfilled, even locally. Indeed, we observe that the order parameter does no longer follow the $\frac{1}{\sqrt{6}}(2,-1,-1)$ behavior associated with the $d_{x^2-y^2}$ order even when the $d_{xy}$-wave is completely suppressed. To completely understand the transition to the smallest members of the harmonic honeycomb series, we instead have to study the point group of the full 3D structure.

%
\section{Hyper- and stripyhoneycomb lattices\label{sec:hypstrip}}
To scrutinize the superconducting order present in the members of the harmonic honeycombs with small distances between twists, we study the two smallest members of the family, the hyper- and stripyhoneycomb lattices. Both structures have been experimentally stabilized in lithium iridate compounds.\cite{Modic2014, Takayama2015}

\begin{figure}[t]
    \includegraphics[width=.3\textwidth]{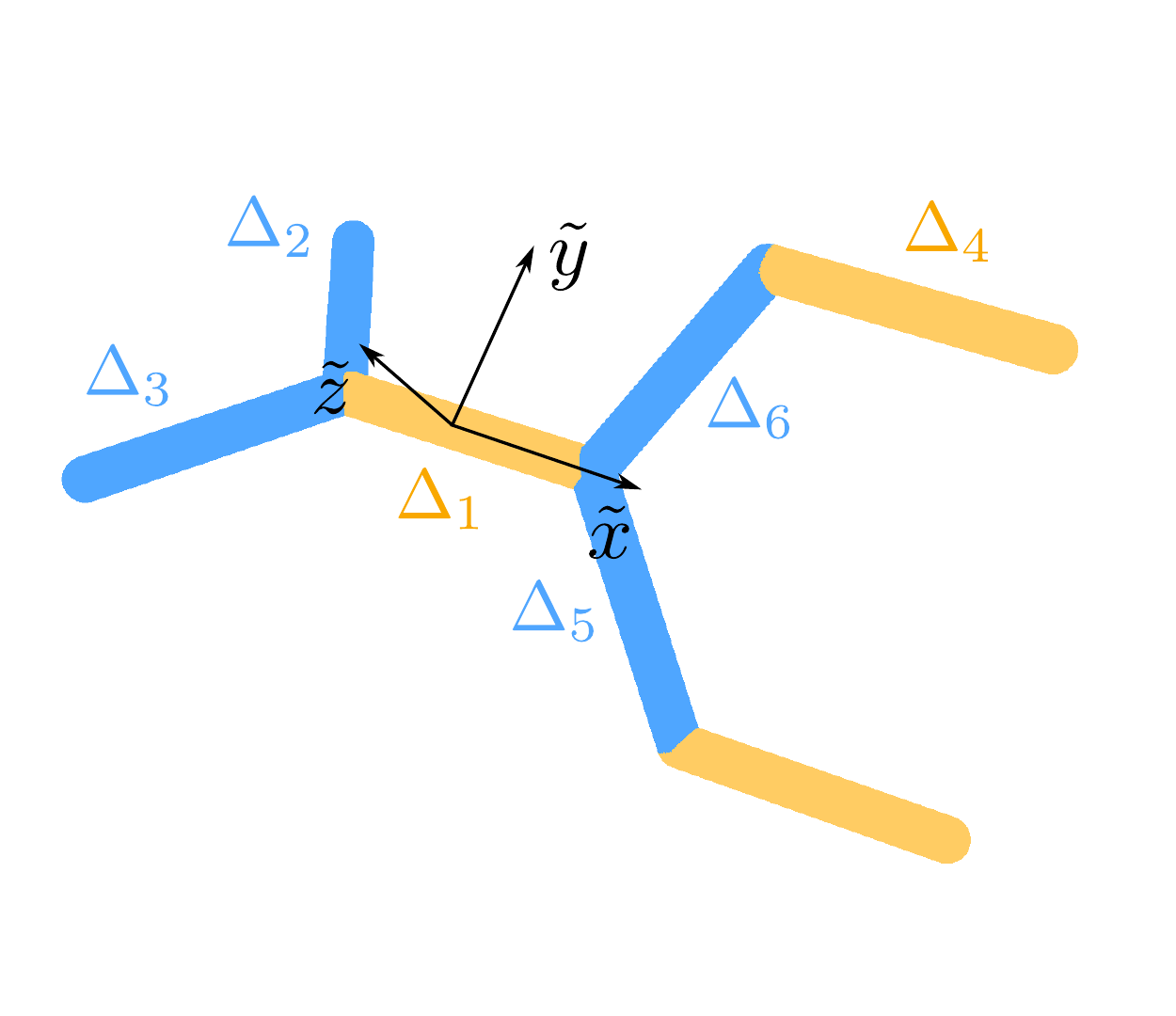}
    \caption{Hyperhoneycomb lattice with its six unique bonds split into two sets. There is no symmetry relating the order parameters on the horizontal bonds $\{\Delta_1,\Delta_4\}$ (yellow) with those on the zigzag bonds $\{\Delta_2,\Delta_3,\Delta_5,\Delta_6\}$ (blue). Black arrows depict the three $C_2$ axes. \label{fig:hypernum}}
\end{figure}

We begin with the hyperhoneycomb lattice, which has the point group $D_{2h}$, with the bond center of one twist bond chosen as the center of symmetry. There are three $C_2$ rotation axes going through this bond as depicted in Fig.~\ref{fig:hypernum}. $\tilde{x}$ points in the same direction as the $x$-direction introduced in Fig.~\ref{fig:lattices}, while the other two rotation axes point into the directions $\vc{a}_2+\vc{a}_2'$ and $\vc{a}_3+\vc{a}_2'$, respectively. For more information about the symmetry of the hyperhoneycomb lattice, see e.g.~Lee et al.\cite{Lee2014}

The hyperhoneycomb lattice has six different bonds, with their own order parameters, which are labeled in Fig.~\ref{fig:hypernum}. In analogy to the honeycomb case, they can be combined into a vector, $\vc{\Delta} = (\Delta_1,\Delta_2,\Delta_3,\Delta_4,\Delta_5,\Delta_6)$. There are no symmetries that relate the twist bonds to the zigzag bonds, so the order parameters split into two different subsets, consisting of the order parameters on the twist bonds $\{\Delta_1,\Delta_4\}$ and those on the zigzag bonds $\{\Delta_2,\Delta_3,\Delta_5,\Delta_6\}$, respectively. The trivial representation then imposes constraints independently for these bonds, resulting in the following basis functions
\begin{subequations}
\begin{align}
\vc{\Delta}_{A_{1g}}^1 & = (1,0,0,1,0,0),\\
\vc{\Delta}_{A_{1g}}^2 & = (0,1,1,0,1,1).
\end{align}
\end{subequations}
Because the two sets are not related by symmetry, there can be different magnitudes and phases for the two subsets. The remaining three even irreducible representations, relevant for spin-singlet pairing, enforce zero amplitudes on horizontal bonds and additional constraints on the zigzag bonds. These  basis functions are
\begin{subequations}
\begin{align}
    		 \vc{\Delta}_{B_{1g}}&= (0,1,-1,0,1,-1),\\
    		 \vc{\Delta}_{B_{2g}}&= (0,1,-1,0,-1,1),\\
    		 \vc{\Delta}_{B_{3g}}&= (0,1,1,0,-1,-1).
\end{align}
\end{subequations}
\begin{figure}[t]
    \includegraphics[width=.4\textwidth]{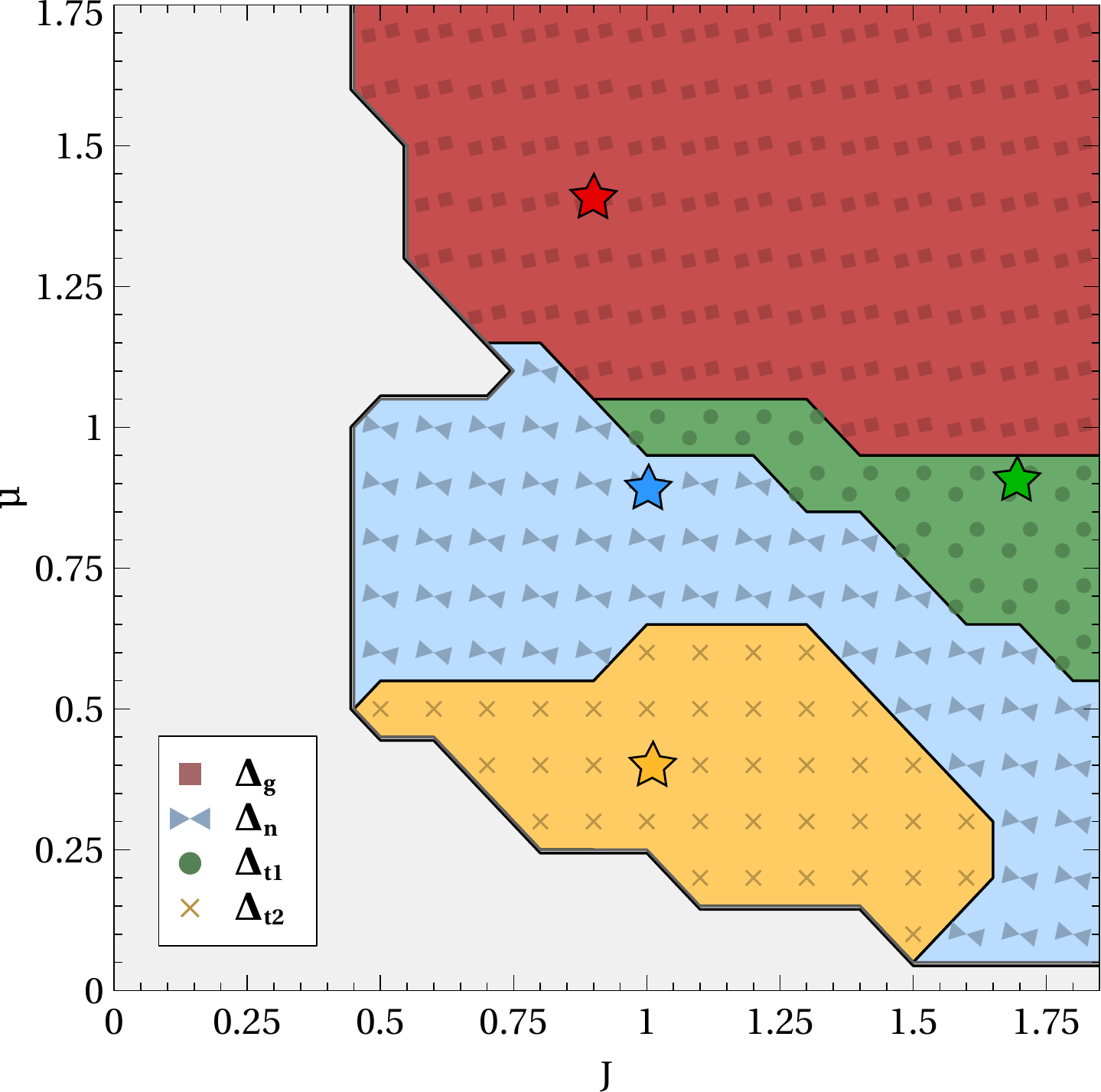}
    \caption{$J$-$\mu$ phase diagram of the hyperhoneycomb lattice at zero temperature. At low doping the system stabilizes in the nodal state (blue region), while for higher doping the system is completely gapped (red region). Between these two regions a completely gapped time-reversal symmetry breaking state is found (green region). In a region at low doping the system favors a time-reversal symmetry breaking state with nodal points (orange region). In the shaded region a classification of the order parameter is not possible within our numerical precision. The stars mark the positions where the DOS in Fig.~\ref{fig:hyperDOS} is obtained. 
    \label{fig:hyperJmu}}
\end{figure}
To see which orders are realized in the hyperhoneycomb lattice, we scan the parameters $J$ and $\mu$ and classify the order parameters in terms of the presented basis functions. Figure~\ref{fig:hyperJmu} shows the resulting phase diagram at zero temperature. The two dominant orders belong to the trivial irreducible representation. At large doping the order parameter can be described by the linear combination
\begin{align}
\vc{\Delta}_{g} &= a \vc{\Delta}_{A_{1g}}^1 + b \vc{\Delta}_{A_{1g}}^2 \nonumber \\
&= (a, b, b, a, b, b), 
\end{align}
where $a$ and $b$ are positive real numbers. The two subsets thus have a different magnitudes, but no relative phase. This state has a fully gapped spectrum as seen in Fig.~\ref{fig:hyperDOS} and we refer to it as the gapped state and use the label $\vc{\Delta}_{g}$. 
At lower doping, another linear combination is stabilized
\begin{align}
\vc{\Delta}_{n} &= a\vc{\Delta}_{A_{1g}}^1 - b \vc{\Delta}_{A_{1g}}^2 \nonumber \\
&= (a, -b, -b, a, -b, -b). 
\end{align}
Again, the magnitude varies between the two sets of bonds, but there is also a relative phase of $\pi$ between them. This sign change leads to the formation of nodal lines in 3D, as evident from the low-energy linear DOS in Fig.~\ref{fig:hyperDOS}. Therefore, we refer to this state as the nodal state and label it $\vc{\Delta}_{n}$.
In a small intermediate region at large interaction strength, the system forms a time-reversal symmetry breaking state belonging to the trivial irreducible representation. It can be written as
\begin{align}
\vc{\Delta}_{t_1} &= a \vc{\Delta}_{A_{1g}}^1 + b \e{\I \phi} \vc{\Delta}_{A_{1g}}^2 \nonumber \\
&= (a, b \e{\I \phi}, b \e{\I \phi}, a, b \e{\I \phi}, b \e{\I \phi}), 
\end{align}
where the relative phase $\phi$ is different from $0$ and $\pi$. This state also completely gaps out the Fermi surface.
At low doping another time-reversal symmetry breaking state develops. It does not belong to only one of the irreducible representations. Instead, the nodal state mixes with the other irreducible representations forming
\begin{align}
\vc{\Delta}_{t_2} &= a \vc{\Delta}_{A_{1g}}^1 - b \vc{\Delta}_{A_{1g}}^2 + \I c \vc{\Delta}_{B_{1/2g}} \nonumber \\
&= (a, -b + \I c, -b - \I c, a, -b \mp \I c, -b \pm \I c ),
\end{align}
which is equivalent to a linear combination of the form $\vc{\Delta}_n + \I c \vc{\Delta}_{B_{1/2g}}$, where $c$ is also real and positive. The time-reversal symmetry breaking imaginary part belongs to either the $B_{1g}$ or the $B_{2g}$ irreducible representations, with the two solutions found to be degenerate within our numerical accuracy. The nodal lines present without the imaginary part are now largely gapped out, with only individual point nodes remaining.

\begin{figure}[t]
    \includegraphics[width=.4\textwidth]{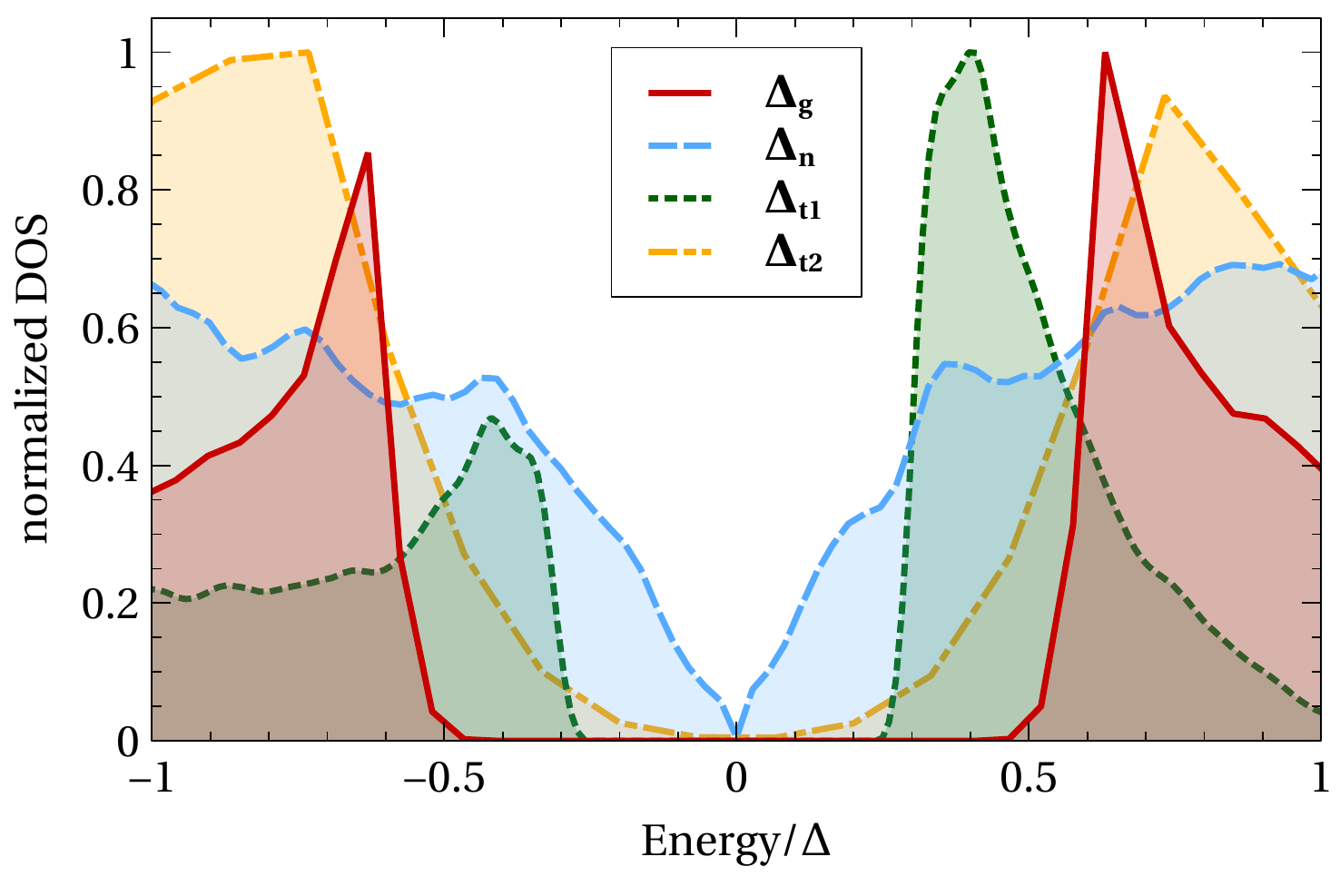}
    \caption{DOS as a function of energy obtained at four different points in the phase diagram, marked with stars in Fig.~\ref{fig:hyperJmu}. All curves are normalized by their maximum value and the energy is rescaled by the magnitude of the order parameter. Each superconducting order displays a very characteristic behavior, ranging from fully gapped, to nodal points and lines. \label{fig:hyperDOS}}
\end{figure}
In Fig.~\ref{fig:hyperDOS} we compare the DOS for all four different superconducting states on the hyperhoneycomb lattice taken at the star marked points in Fig.~\ref{fig:hyperJmu}. For the state at high doping, the fully opened gap is clearly visible. 
A linear relationship between DOS and energy around zero is characteristic of line nodes in 3D, while when the line nodes are gapped out, the DOS has a quadratic energy dependence. The time-reversal symmetry breaking state of the trivial irreducible representation also completely gaps the spectrum.

\begin{figure}[t]
    \includegraphics[width=.4\textwidth]{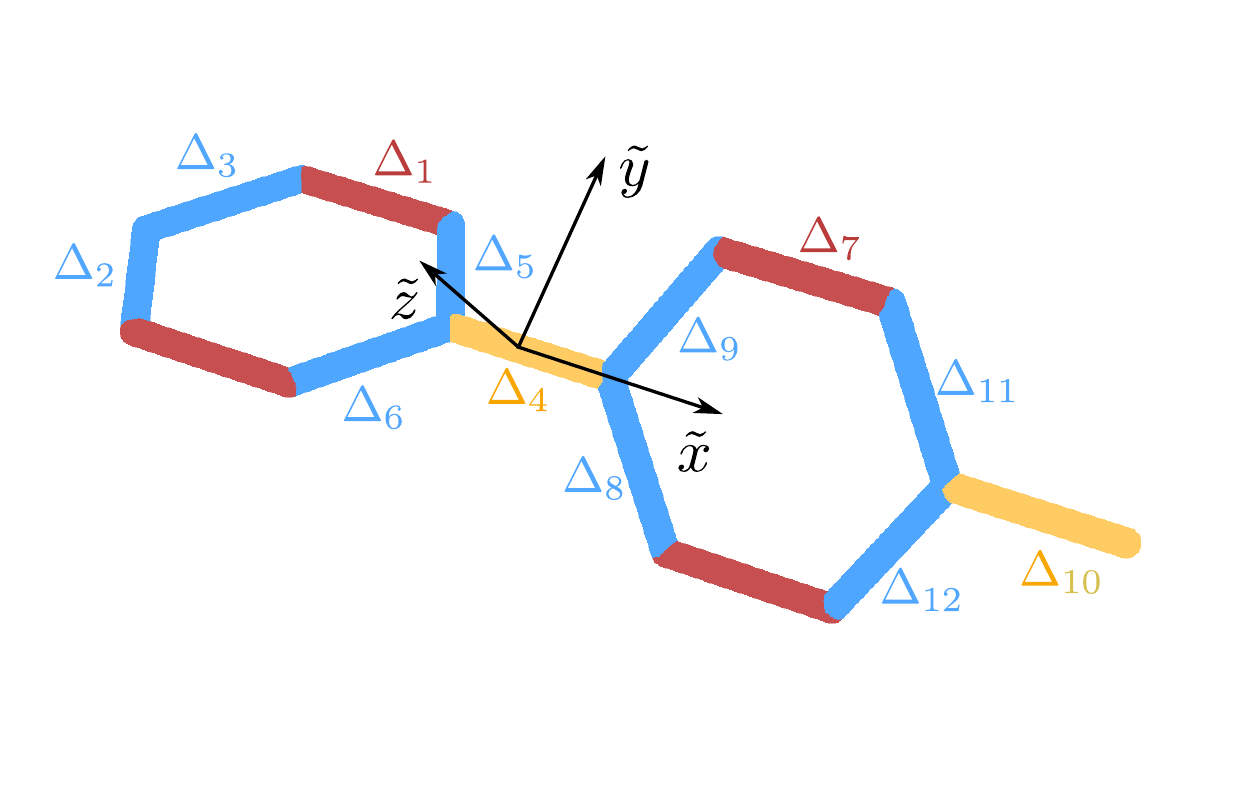}
    \caption{Stripyhoneycomb lattice with its 12 unique nearest neighbor bonds split into three sets which are not related by symmetry. Black arrows depict the three $C_2$ axes. \label{fig:stripynum}}
\end{figure}
A similar study can be performed for the stripyhoneycomb lattice, which also has the point group $D_{2h}$. The center of symmetry is in the bond center of one of the two twist bonds, see Figure~\ref{fig:stripynum}.
The $\vc{\Delta}$ vector here contains twelve individual bond order parameters, one for each bond. As Fig.~\ref{fig:stripynum} shows, another group of (horizontal) bonds is added to the lattice, $\{\Delta_1, \Delta_7\}$. Also, there are now four more zigzag bonds, $\{\Delta_2, \Delta_3, \Delta_5, \Delta_6, \Delta_8,\Delta_9, \Delta_{11}, \Delta_{12}\}$, while there are still only two twist bonds, $\{\Delta_4, \Delta_{10} \}$. Again, there is no symmetry that maps members of one group onto members of another. Thus, in the trivial irreducible representation, three basis functions arise, one for each group of bonds
		\begin{subequations}	    
	    \begin{align}
		  \vc{\Delta}_{A_{1g}}^1 &= (1,0,0,0,0,0,1,0,0,0,0,0),\\
		  \vc{\Delta}_{A_{1g}}^2 &= (0,0,0,1,0,0,0,0,0,1,0,0),\\
    		  \vc{\Delta}_{A_{1g}}^3 &= (0,1,1,0,1,1,0,1,1,0,1,1).
        \end{align}
        \end{subequations}
As in the hyperhoneycomb lattice, the magnitude and relative phases between the sets are not restricted by symmetry operations of the lattice. The basis functions of the other three irreducible representations again only involve the zigzag bonds and introduce different signs on these bonds
		\begin{subequations}	    
	    \begin{align}
		  \vc{\Delta}_{B_{1g}} &= (0,1,-1,0,1,-1,0,1,-1,0,1,-1),\\
		  \vc{\Delta}_{B_{2g}} &= (0,1,-1,0,1,-1,0,-1,1,0,-1,1),\\
    		  \vc{\Delta}_{B_{3g}} &= (0,1,1,0,1,1,0,-1,-1,0,-1,-1).
        \end{align}
        \end{subequations}
        
\begin{figure}[t]
    \includegraphics[width=.4\textwidth]{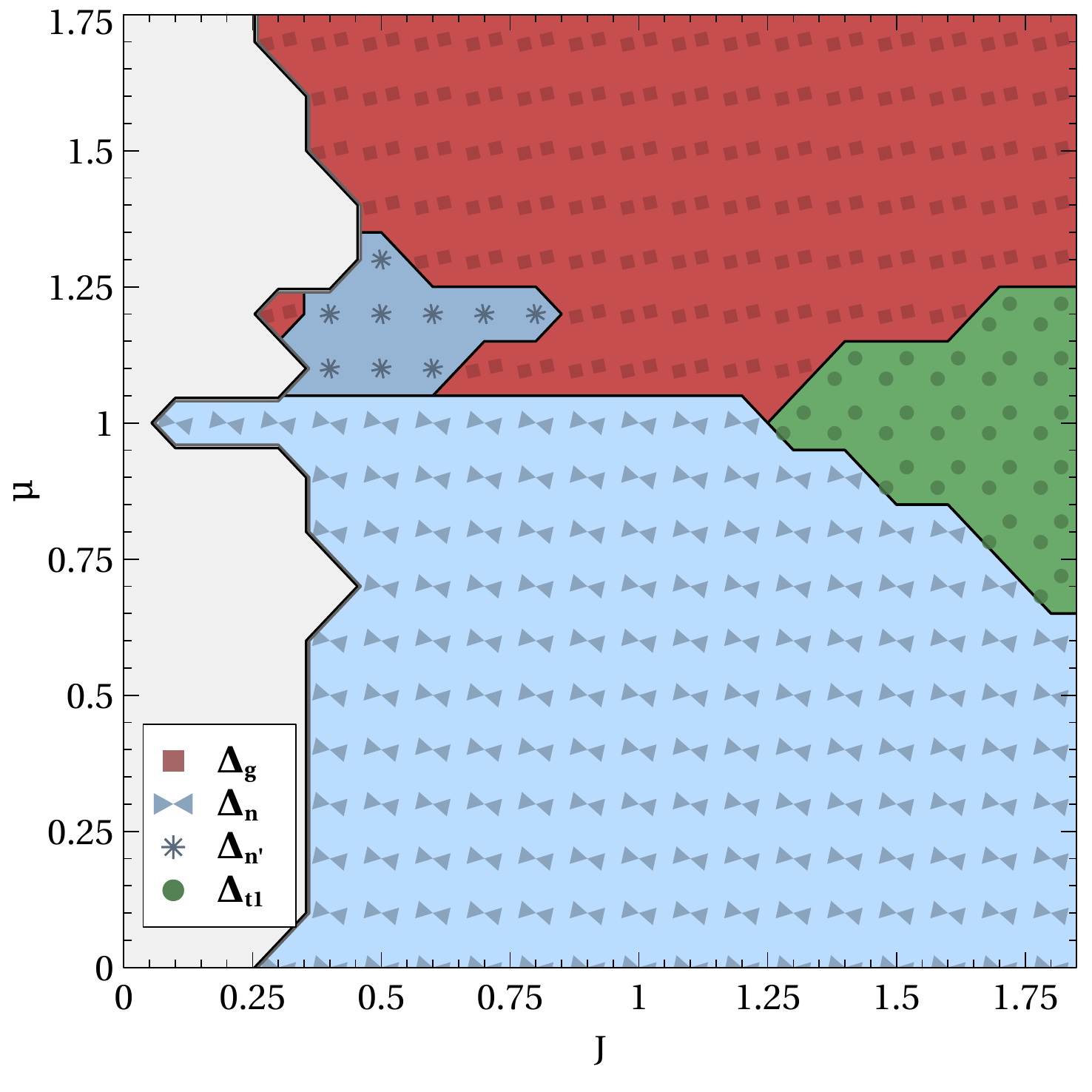}
    \caption{$J$-$\mu$ phase diagram of the stripyhoneycomb lattice at zero temperature. Two different nodal states are found for low doping (blue regions). At high doping the system is fully gapped (red region) and in an intermediate region at high interaction strengths the order parameter breaks time-reversal symmetry (green region). In the shaded region a classification of the order parameter is not possible within our numerical precision.  \label{fig:stripyJmu}}
\end{figure}  
In Fig.~~\ref{fig:stripyJmu} we plot the $J-\mu$ phase diagram for the stripyhoneycomb lattice at zero temperature. We identify four distinct superconducting phases, three of which are directly related to the orders observed in the hyperhoneycomb lattice. At large doping, a fully gapped solution of the form
\begin{align}
\vc{\Delta}_{g} &= a \vc{\Delta}_{A_{1g}}^1 + b \vc{\Delta}_{A_{1g}}^2 + c \vc{\Delta}_{A_{1g}}^3
\end{align}
is stabilized, where again $a,b, c$ are real and positive constants. When the doping level is lower, there are two different nodal states. One of them has a phase difference of $\pi$ between horizontal and zigzag bonds and can be expressed as 
\begin{align}
\vc{\Delta}_{n} &= a \vc{\Delta}_{A_{1g}}^1 + b \vc{\Delta}_{A_{1g}}^2 - c \vc{\Delta}_{A_{1g}}^3. 
\end{align}
The other one features a $\pi$ phase difference between the horizontal and the twist bonds, which corresponds to the linear combination
\begin{align}
\vc{\Delta}_{n'} &= a \vc{\Delta}_{A_{1g}}^1 - b \vc{\Delta}_{A_{1g}}^2 + c \vc{\Delta}_{A_{1g}}^3. 
\end{align}
As in the hyperhoneycomb lattice, there is also an intermediate region at high interaction strengths, where the superconducting order breaks time-reversal symmetry by forming a linear combination of the form
\begin{align}
\vc{\Delta}_{t_1} &= a \vc{\Delta}_{A_{1g}}^1 + b \vc{\Delta}_{A_{1g}}^2 + c \e{\I \phi} \vc{\Delta}_{A_{1g}}^3,
\end{align}
where $\phi$ is different from $0$ and $\pi$. This state completely gaps the Fermi surface.
%
\section{Trends for the harmonic honeycomb lattices \label{sec:connection}}

With the knowledge of the superconducting states appearing in the hyper- and stripyhoneycomb lattice, it is possible to draw conclusions about the superconducting order parameter for other smaller members of the harmonic honeycomb series. Several trends can be observed. First of all, the point group of all members of the series is $D_{2h}$, which allows us to generalize the results quite easily. Secondly, as the step from hyper- to stripyhoneycomb shows, increasing the distance between twists will add further sets of bonds, either of the zigzag or of the horizontal kind, that are not related by symmetry to any of the previous bonds. The magnitude of the order parameter in each set of bonds is thus free and not related to the other sets. Within each set, we can conclude that the trivial irreducible representation enforces the same order parameter on each bond, but does not dictate the phases between the different set of bonds. When the relative phase between horizontal and zigzag bonds equals $\pi$, nodal lines appear in the spectrum. This state is expected to be the most stable state at low doping. For higher doping, we expect a fully gapped state to form, which requires the same phase on all sets of bonds. Other irreducible representations are classified by order parameters alternating in sign along the zigzag bonds. 

When the distance between twists is large, there is locally an approximative six-fold symmetry. Then the superconducting order can be classified with respect also to the honeycomb $D_{6h}$ point group and the states found originally in the hyper- and stripyhoneycomb can be related to those of the honeycomb lattice. For the gapped state $\vc{\Delta}_g$ observed at large doping levels $\mu  > 1$, the additional symmetry fixes the magnitude of the order parameter on the three bonds around each site to the same value. This is equivalent to the extended $s$-wave state, which is also obtained at approximately the same high doping levels in the honeycomb lattice.\cite{Lothman2014} For lower doping levels the nodal state $\vc{\Delta}_n =(a, -b, -b, a, -b, -b)$, with a relative phase of $\pi$ between horizontal and zigzag bonds, dominates the phase diagram for both the hyper- and stripyhoneycomb lattices. An exact local symmetry restricts $a = 2 b$, which is also the spatial basis for the $d_{x^2-y^2}$-wave of the $D_{6h}$ point group. This explains the results in Fig.~\ref{fig:dvsL}, where the overlap with the $d_{x^2-y^2}$-wave never quite reaches one due to this local symmetry only being approximately valid. The sign-changing zigzag bond orders of the other irreducible representations can be be related to the $d_{xy}$ state, since it has the vector order parameter $\vc{\Delta}_{d_{xy}} \sim (0,1,-1)$, which changes sign on the zigzag bonds. The time-reversal symmetry breaking order with nodal points, $\vc{\Delta}_{t_2}$, found at low doping in the hyperhoneycomb lattice (but not in the stripyhoneycomb) therefore corresponds to the $d \pm \I d'$ state observed throughout the phase diagram at low to moderate doping in the honeycomb lattice. In summary, we thus find that the superconducting state on the harmonic honeycomb lattices at low doping levels is related to the $d_{x^2-y^2}$-wave state of the honeycomb lattice due to the pair breaking effects of the zigzag edge on the $d_{xy}$-state. Still, in some limited regions it is possible to stabilize the equivalent of the $d+id'$ state, despite the two $d$-wave solutions belonging to different irreducible representations for all harmonic honeycomb lattices.

Another feature of the phase diagrams that also appears in the honeycomb lattice is the change of order at roughly $\mu = 1$, from $s$-wave like to $d$-wave like. For all the members of the harmonic honeycomb lattice we have studied, some kind of feature (either a shoulder or a clear peak) appears at this doping level in the normal state DOS, very similar to the van Hove singularity found at $\mu = 1$ in the honeycomb lattice. Finally, the observed suppression of the order parameter on the bonds at the twist itself is enabled by the fact that these bonds form one group that is not related by symmetry to the any other bonds.
%
\section{Conclusions\label{sec:conclusion}}

In this work we have studied the possible spin-singlet superconducting states in harmonic honeycomb materials upon doping an  Heisenberg antiferromagnet ground state.  
We first studied an isolated twist between two separate honeycomb lattice regions, rotated out-of-plane relative to each other. Here we found that the two regions of honeycomb structure separated by the twist independently stabilize the chiral $d \pm \I d'$-wave solution at low to intermediate doping. The twist effectively acts as an open edge in the system, as it does not constrain the relative chirality between the two regions and it hosts a total of four edge states, two chiral states from each side of the twist. However, the twist still offers some coupling between the two regions as is evident from the edge states dispersing along both directions in reciprocal space, forming 2D edge bands. At very high doping levels ($\mu \gtrsim 1$), we instead found an extended $s$-wave state on both sides of the twist.  

We then introduced periodic boundary conditions by which we could decrease the region size between the twists. We then found that the degeneracy of the two $d$-wave states is lifted since the zigzag edge is pair breaking for the $d_{xy}$-wave state.
When the distance between twists become too small, the $D_{6h}$ symmetry of the honeycomb lattice is no longer preserved even locally and the superconducting state has to be classified in terms of the $D_{2h}$ point group of the harmonic honeycomb lattice series.
Because there are no symmetries relating certain sets of bonds to each other in the harmonic honeycomb lattices, the bond order parameters split into several separate sets. This leads to the formation of superconducting states with nodal lines belonging to the trivial irreducible representation at low to intermediate doping in both the hyper- and stripyhoneycomb lattices. This state is the natural evolution of the $d_{x^2-y^2}$ state in the honeycomb lattice. 
The nodal lines can in fact be shown to be topological protected, as will be discussed elsewhere.\cite{Bouhon2016}
In parts of the low-doping regime of the hyperhoneycomb phase diagram we also found a time-reversal symmetry breaking state that partly gaps out the nodal lines into nodal points. This state belongs to different irreducible representations, but is an extension of the honeycomb $d+id'$-wave state.  
At very high doping we find a fully gapped state corresponding to an extended-$s$-wave state. In between the low-moderate and high doping regimes there is also an additional fully gapped time-reversal symmetry breaking state generated from the trivial representation, with no equivalence in the 2D honeycomb lattice.
In aggregate these results display the evolution from a chiral and fully gapped $d+id'$-wave superconducting state in the 2D honeycomb lattice to a superconducting state with nodal lines in the small members of the 3D harmonic honeycomb lattices.

\begin{acknowledgments}
We are grateful to M.~Hermanns and K.~Le Hur for discussions. This work was supported by the Swedish Research Council (Vetenskapsr\aa det), the Swedish Foundation for Strategic Research (SSF), the G\"{o}ran Gustafsson Foundation, and the Wallenberg Academy Fellows program.
\end{acknowledgments}

\bibliographystyle{apsrevmy}

\end{document}